\documentclass[aps,prb,twocolumn,showpacs]{revtex4}

\usepackage{graphicx}
\usepackage{psfrag}
\usepackage{dcolumn}
\usepackage{bm}
\usepackage{amssymb}

\def\kp{${\bf k}\cdot{\bf p}$}
\def\GaMnAs{Ga$_{1-x}$Mn$_x$As}

\begin{document}
\title{Disorder Enhanced Spin Polarization in Diluted Magnetic Semiconductors}

\author{Byounghak Lee}
\email{bhlee@lbl.gov}
\affiliation{Computational Research Division, Lawrence Berkeley National Laboratory, Berkeley, California 94720}
\author{Xavier Cartoix\`a}
\affiliation{Department d'Enginyeria Electr\`onica, Universitat Aut\`onoma de Barcelona, 08193 Bellaterra, Spain}
\author{Nandini Trivedi}
\affiliation{Department of Physics, The Ohio State University, Columbus, Ohio 43210}
\author{Richard M. Martin}
\affiliation{Department of Physics, University of Illinois at Urbana-Champaign, Urbana, Illinois 61801}

\date{\today}

\begin{abstract}
We present a theoretical study of diluted magnetic semiconductors that includes spin-orbit coupling within a realistic host band structure and treats explicitly the effects of disorder due to randomly substituted Mn ions.
While spin-orbit coupling reduces the spin polarization by mixing different spin states in the valence bands, we find that disorder from Mn ions enhances the spin polarization due to formation of ferromagnetic impurity clusters and impurity bound states.
The disorder leads to large effects on the hole carriers which form impurity bands as well as hybridizing with the valence band.
For Mn doping $0.01 \lesssim x \lesssim 0.04$, the system is metallic with a large effective mass and low mobility.
\end{abstract}


\maketitle

%%%%%%%%%%%%%%%%%%%%%%%%%%%%%%%%%%%%%%%%%%%%%%%%%%%

Ferromagnetic semiconductors with high transition temperatures and large polarization are important for spintronics applications.
The III-V diluted magnetic semiconductors (DMS) showing ferromagnetism with Curie temperatures $T_c~\sim 170 K$\cite{Ohno1998Making-Nonmagne, Wang2005Proceedings-of-} hold considerable promise.\cite{2002Semiconductor-S}
It is generally agreed that ferromagnetism between the magnetic impurities is mediated via the interaction with itinerant carriers;\cite{Jungwirth2006Theory-of-ferro} however, a comprehensive theoretical understanding has been hampered by the complexity arising from the interplay of electron-local moment interaction, disorder from the doped Mn ions, and spin-orbit (SO) effects.

Theoretical predictions based on the density functional theory (DFT)\cite{Sanvito2001First-principle,Jain2001Electronic-stru,Kulatov2002Electronic-stru,Zhao2004Comparison-of-p} have predicted half-metallicity of \GaMnAs~ and other DMS.
These calculations are performed with a supercell containing a small number of atoms, often with only one Mn ion in the simulation cell, ruling out any disorder effects.
Other methods, such as the KKR-CPA-LDA method\cite{Akai1998Ferromagnetism-}(Korringa-Kohn-Rostoker method with coherent potential approximation and local density approximation of density functional method), include disorder effects only within an effective medium approximation.
In addition, most of DFT calculations are performed without including SO coupling and, as a result, overestimate the spin-polarization of carriers.
Other theoretical approaches based on single-band tight-binding method~\cite{Berciu2001Effects-of-Diso,Timm2002Correlated-Defe,Alvarez2002Phase-Diagram-o} can treat disorder effects properly but do not include SO coupling effects.
Just as the DFT method, these approaches tend to overestimate the spin-polarization.
Multiband methods based on \kp~ approximation\cite{Jungwirth1999Interlayer-coup,Dietl2000Zener-model-des} includes the SO coupling effects and realistic band structure for the host but the disorder effects are completely ignored.

In this Letter, we address the interplay of disorder and SO effects on the electronic structure and the spin-polarization of DMS by using a realistic multiband tight-binding model.
The magnetic coupling between the localized Mn substitutional impurity electrons and the itinerant host electrons is included via an exchange parameter.
We find the hole states have dual characteristics with weight in an impurity band as well as in the valence band of the host GaAs, resolving the controversial issue of the location of the holes donated by Mn.
The density of states for low impurity concentration ($\sim$ 1\% Mn) shows deep in-gap localized states at Mn sites.
Even at low doping, strong hybridization with the valence band states pulls considerable weight in the gap region which remains separated from the impurity band.
For higher dopings of about $4\%$ there is some mixing between impurity and valence band states, but the impurity states still retain their distinct characteristics and are more localized than the valence band states..
These states around the impurity band have a larger effective mass and lower mobility than parent semiconductor valence band states, and have recently been observed through the optical transitions between valence and impurity states in midinfrared region~\cite{Burch2006-Impurity-Band-}.

Another very surprising feature of our results is that spin polarization is larger in disordered systems compared with ordered systems for the same Mn concentration, due to the formation of strongly polarized bound states and Mn clusters.
Similar enhancement of the Curie temperature was found within a single-band tight-binding model for DMS.\cite{Berciu2001Effects-of-Diso}

%%%%%%%%%%%%%%%%%%%%%%%%%%%%%%%%%%%%%%%%%%%%%%%%%%%
{\em Model:}
We start with the $sp$-$d$ exchange model $H=H_0 + H_{ex}$, containing the Hamiltonian of the host semiconductor in the clean limit $H_0$ and the effective exchange coupling between the localized and itinerant spins $H_{ex}$.\cite{Kossut1976ELECTRON-TRANSP}
The electronic structure of GaAs is described by the Effective Bond Orbital Model (EBOM) developed by Chang.\cite{Chang1988Bond-orbital-mo}
The effect of the magnetic impurities is described by
\begin{equation}
H_{ex} = - \sum_{I,n,m} J_{n m} {\bf S}_I \cdot ( c_{I n}^{\dagger} {\bm \sigma}_{n m} c_{I m} ) \;,
\label{eq:kineticexchange}
\end{equation}
where ${\bm \sigma}$ is the Pauli spin matrix, and ${\bf S}_I$ are the local spin moments arising from the impurities $I$, which we treat as classical spins of magnitude $S$=5/2.
$n$ and $m$ are the combined indices of the spin and the bond orbital, i.e., $n=\{\tau, \alpha\}$, where in the current 8-band model $\tau$=$\{\uparrow, \downarrow\}$ and $\alpha$=$\{s,p_x,p_y,p_z\}$.
The impurities are placed at the EBOM fcc lattice sites and the matrix elements are calculated in the effective bond basis.
The exchange couplings capture the hybridization of impurity states with itinerant electron states as well as the interaction between electrons mediated by impurity sites.
Given that Mn atoms in GaAs are substitutional impurities with site symmetry of the Ga sites,\cite{Shioda1998Local-structure} it can be readily shown that the exchange coupling does not mix different orbital states; $J_{nm}=J_{\tau \alpha, \tau^\prime \alpha^\prime} \delta_{\tau \tau^\prime} \delta_{\alpha \alpha^\prime}$.
By comparing the single impurity bound state energy with band edge emission measurements~\cite{Lee1964Edge-emission-i} of 113 meV, we determine the exchange coupling between the valence band holes and Mn 3d electrons, $J_{pd} \equiv J_{\tau p, \tau p}$ = -2.48 eV.
The ferromagnetic exchange coupling of conduction electrons with Mn 3d electrons, $J_{sd} \equiv J_{\tau s, \tau s}$, are known to be very weak, generally of order 0.1 eV.\cite{Szczytko1996The-sp-d-exchan}
We find our results are insensitive to this coupling and hereafter we set $J_{sd}$ = 0.1 eV.

\begin{figure}[b]
\centerline{ \includegraphics[width=0.7\columnwidth]{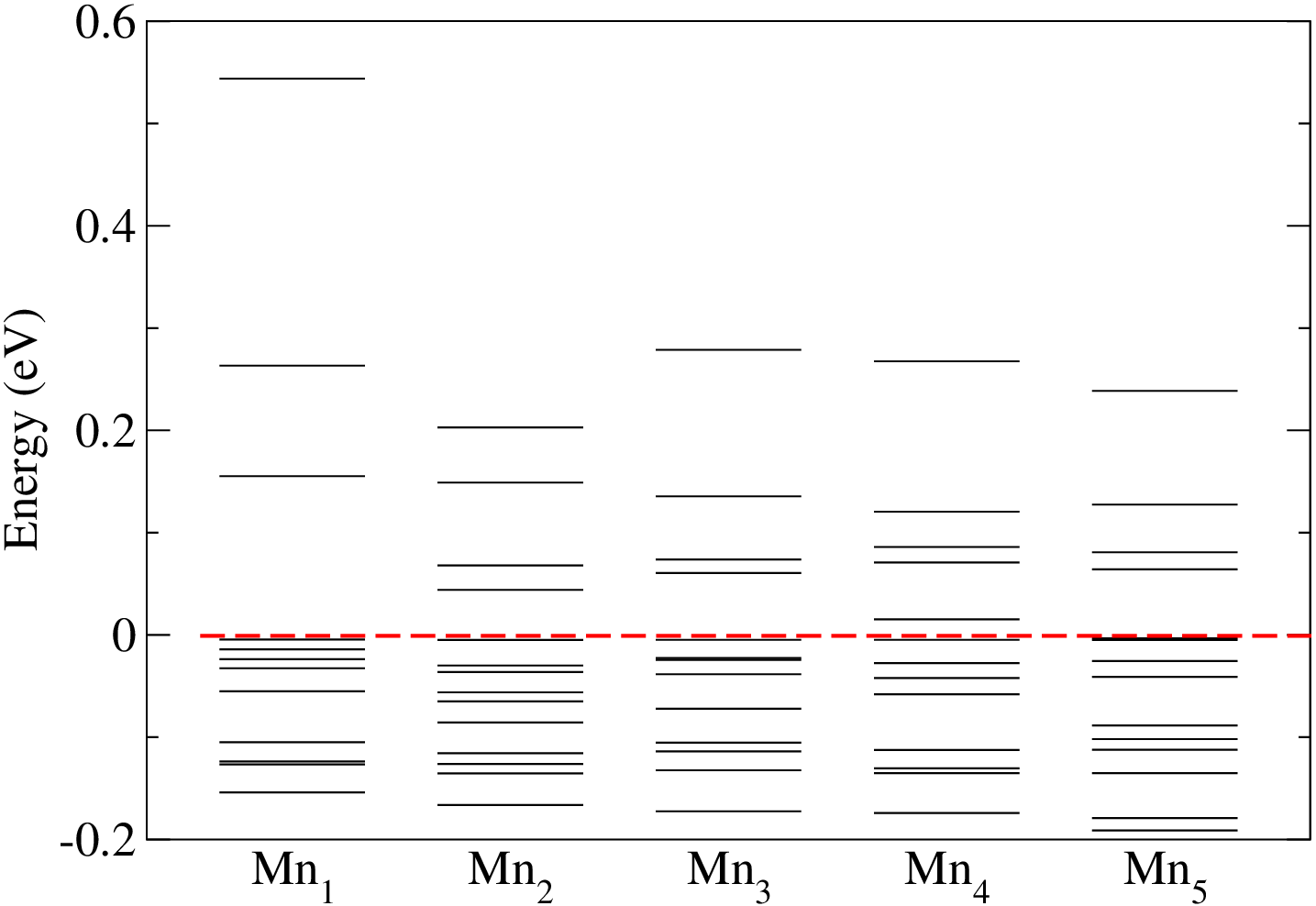} }
\caption{
Binding energy of carriers to Mn-Mn pairs.
Mn$_i$ indicates the $i$th nearest-neighbor Mn pair configuration.
The horizontal dashed line indicates the position of the valence band maximum of host GaAs.
}
\label{fig:TwoImp}
\end{figure}

The advantage of our EBOM based $sp$-$d$ kinetic exchange model is that both the realistic band structure of host materials, including the SO effects, and the disorder effects are treated accurately and on an equal footing.
EBOM not only reproduces \kp\ model results at the zone center but also yields much better band structure away from the zone center.
The disorder effects due to the random position of Mn ions are captured exactly for a given realization of the Mn ion configuration.

%%%%%%%%%%%%%%%%%%%%%%%%%%%%%%%%%%%%%%%%%%%%%%%%%%%
{\em Bound states of Mn pairs:}
Before we look into many-Mn ion configurations, we investigate the carrier states of a Mn ion pair, simulated using a supercell containing 586 GaAs primitive cells and two Mn ions, shown in Fig.~\ref{fig:TwoImp}.
Note that the carrier states of a Mn ion pair are very strongly bound, located well inside the GaAs band gap region, i.e. energy region above 0 eV in Fig.~\ref{fig:TwoImp}.
The nearest neighbor Mn impurity pair forms the deepest bound state with an energy $\approx$ 0.55 eV above the GaAs valence band maximum.
The spin-resolved local density of states (DOS) of Mn pairs shows that the bound states are completeley spin-polarized.
Bound states of large MnAs clusters of more than two Mn ions generally induce even deeper bound states (not shown here).
The non-monotonic distance dependence of the pair bound states indicates a strong anisotropy of the ferromagnetic coupling due to the zinc-blende crystal symmetry.

We next analyze the effect of a finite concentration of impurities.

%%%%%%%%%%%%%%%%%%%%%%%%%%%%%%%%%%%%%%%%%%%%%%%%%%
\begin{figure}[b]
 \centerline{\includegraphics[width=0.9\columnwidth]{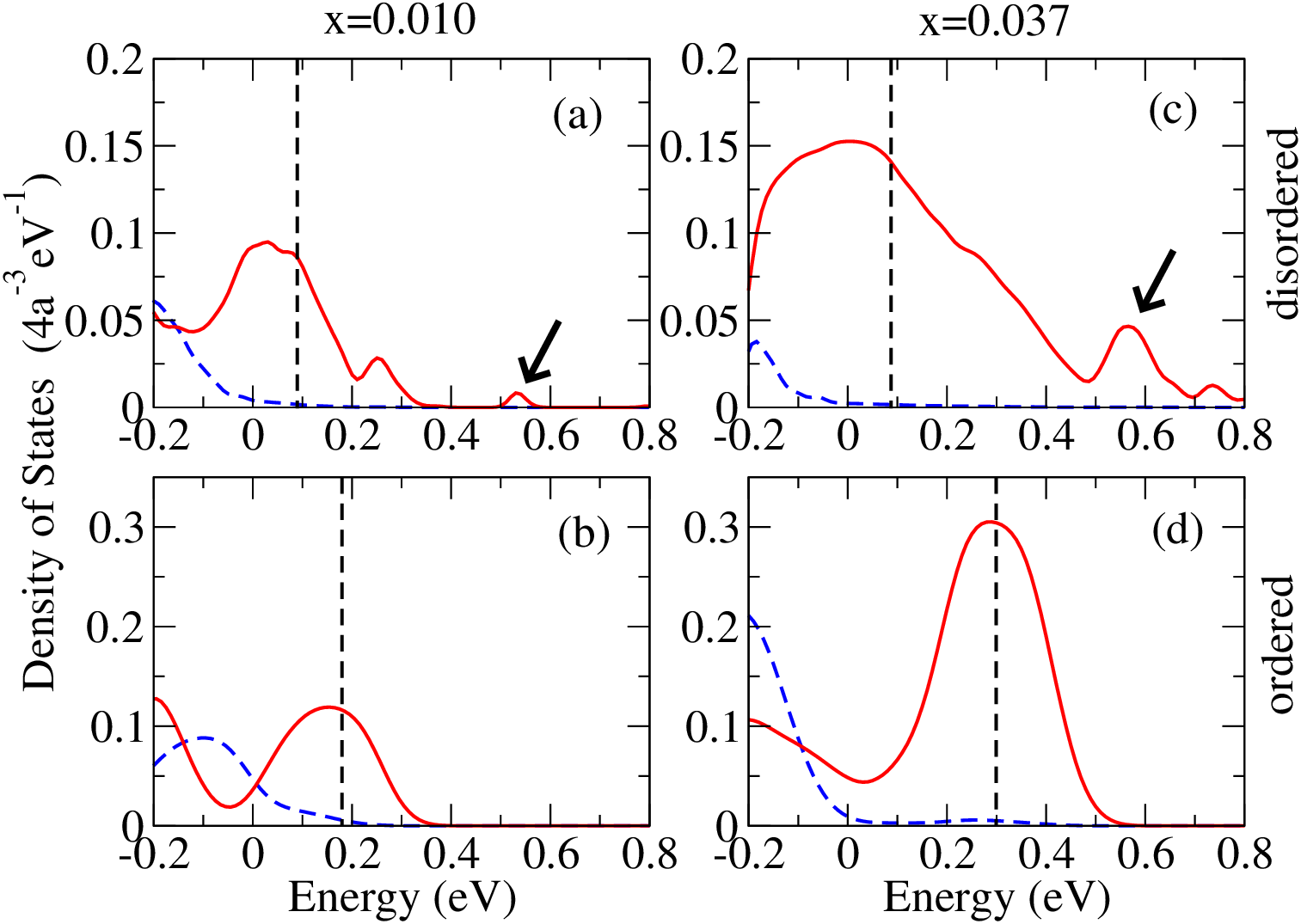}}
 \caption{
 Density of States (DOS) of \GaMnAs.
(a) and (c): EBOM with randomly placed Mn ions.
(b) and (d): EBOM with one Mn ion in a supercell.
(a) and (b) are for $x=0.010$ and (c) and (d) are for $x=0.037$.
Solid (red) and dashed (blue) curves correspond to the majority and minority spin DOS.
100 disorder configurations were used for the disorder average.
Energies are relative to the host GaAs valence band maximum energy.
The dashed vertical lines indicate the position of the chemical potential for uncompensated charge density.\cite{footnote1}
Notice that magnetic impurities generate a large DOS in the GaAs band gap region [0, 1.52] eV due to strong hybridization with the valence band.
At both dopings there is a peak in the DOS at $\approx$ 0.55 eV (indicated by arrows in (a) and (c)) originating from the bound states of nearest neighbor Mn ion pairs.
 }
\label{fig:dos}
\end{figure}

{\em Disorder effects:}
Randomly substituted Mn ions are included explicitly by diagonalizing the Hamiltonian $H$ and evaluating the doped hole states for a given local spin configuration. Properties, such as the local DOS and conductivity are calculated for a given Mn configuration are then averaged over a large number of Mn ion configurations.
We assume a collinear magnetic ground state where all localized magnetic ion spins are aligned along the magnetic easy axis along $\langle 100 \rangle$ direction.

Fig.~\ref{fig:dos} (a) and (c) show that random locations of the Mn impurities generate long tails and a peak around the bound state energy of Mn pairs ($\approx$ 0.55 eV) in the GaAs gap region [0, 1.52] eV.
In supercell calculations with the same impurity concentration, Fig.~\ref{fig:dos} (b) and (d), the DOS has large
weight in the GaAs gap region but does not show the long tails or the extra peak, seen in disordered cases.
The impurity band around 0.55 eV is separated from the main bands at low doping concentrations of $x$ = 0.01.
At $x$ = 0.037, the impurity induced states still show a strong peak but get more strongly hybridized with the valence bands.

The deep in-gap states in disordered impurity configuration originate from the bonding states of neighbor impurity pairs.
The position of the largest impurity band DOS, indicated by arrows in Fig.~\ref{fig:dos} (a) and (c), corresponds to the deepest bound states of nearest neighboring Mn pairs in Fig.~\ref{fig:TwoImp}, $\approx$ 0.55 eV.
The deeper states at $\sim$ 0.7 eV in Fig.~\ref{fig:dos} (c) at $x$=0.037 comes from the bound states of cluster of more than two Mn ions.
In marked contrast, calculations based on effective medium theories fail to capture the presence of clusters.
In coherent potential approximation, for example, the impurity-originated states give rise to the impurity band spectral weight near the single impurity bound states, which is $\approx$ 110 meV in \GaMnAs~, but the long tails deep inside the band gap region are not found.

\begin{figure}[b]
 \centerline{\includegraphics[width=0.9\columnwidth]{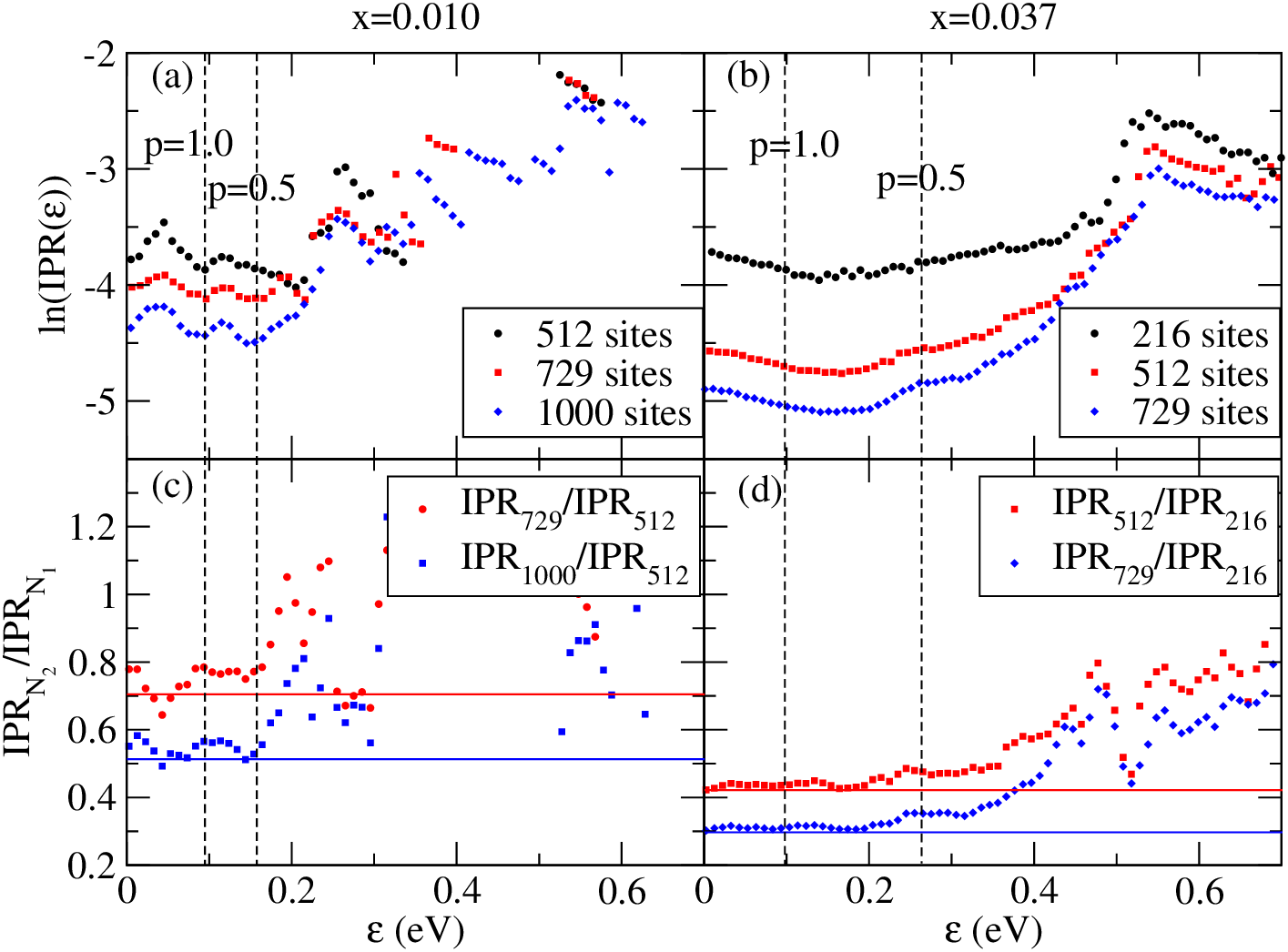}}
 \caption{
 Inverse participation ratio (IPR) of \GaMnAs~.
 Left panels (a) and (c), are for $x$ = 0.010 and right panels (b) and (d), are for $x$ = 0.037.
 (a) and (b): $IPR(\epsilon)$ plotted in logarithmic scale as a function of the carrier energy, $\epsilon$.
 (c) and (d): scaling of $IPR_N$, where $N$ is the size of the system.
 The horizontal lines show the $IPR_{N_2}/IPR_{N_1} \propto N_1/N_2$, expected for extended wavefunctions.
 The energies are relative to the GaAs valence band maximum located at 0 eV.
 The dashed vertical lines denotes the position of the chemical potential at given charge fraction $p$ = (hole density/Mn density).
 Strong size dependence indicates extended states, whereas localized states show little size dependence.
 For each system size, the participation ratio was averaged over 100 disorder realizations.
}
 \label{fig:participation}
\end{figure}

The presence of in-gap states is of great importance for transport and optical properties as they influence the inter- and intra-band transition.
To study the nature of the impurity bands, we calculate the inverse participation ratio,\cite{Edwards1972Numerical-studi}
$ IPR(\epsilon) = \int d{\bf r} |\psi_\epsilon ({\bf r})|^4$,
where $\epsilon$ is the eigenenergy of the Hamiltonian corresponding to the wavefunction $\psi_\epsilon({\bf r})$.
For extended states, $IPR(\epsilon)$ scales with the system size as $\sim 1/V$, where $V$ is the volume of the system whereas for localized states, $IPR(\epsilon)$ is independent of the system size.

Fig.~\ref{fig:participation} shows that at impurity doping concentrations $x$=0.037 the carrier states at the Fermi energy are extended; hole states with energy $\lesssim$ 0.25 eV scale nearly linearly with the size of the system, indicating that carriers are delocalized.
When the carriers are compensated as much as 50\% by other defects, such as As antisites, the wavefunctions are not completely extended.
This localization tendency might explain the heavier hole effective mass in recent midinfrared measurements.\cite{Burch2006-Impurity-Band-}
For $x$=0.01, on the other hand, the states at Fermi level are not completely extended even without charge compensation.
For both impurity concentration, the deep in-gap states are strongly localized.
Our IPR study indicates that the charge compensation due to other defects, such as As$_{\text{Ga}}$ antisites, could increase the hole effective mass and lower the hole mobility if the chemical potential shifted significantly.

%%%%%%%%%%%%%%%%%%%%%%%%%%%%%%%%%%%%%%%%%%%%%%%%%%
\begin{figure}[b]
\centerline{ \includegraphics[width=0.9\columnwidth]{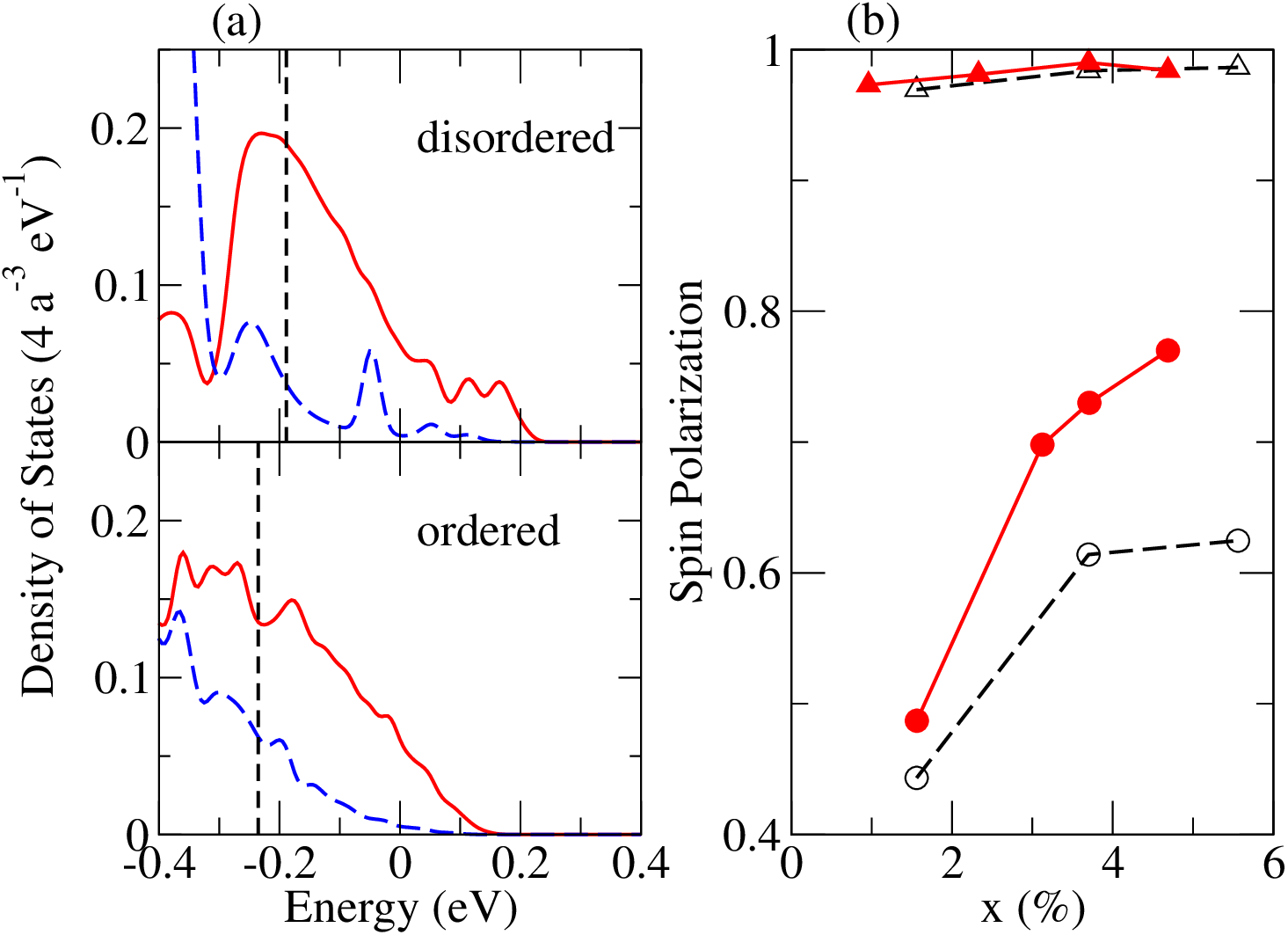} }
\caption{
Disorder effects on spin polarization.
(a) Density of states (DOS) of GaAs with the magnetic impurity concentration $x$ = 0.037 and the exchange coupling $J_{pd}$ = -1.5 eV.
Upper panel: disordered, Lower panel: ordered array.
Red(solid) and blue (dashed) lines denote majority and minority spin DOS, respectively.
The energy is relative to the undoped GaAs valence band maximum.
The vertical lines are the position of the chemical potential.
The hole states with energy above the chemical potential are occupied.
(b) The carrier spin-polarization as a function of impurity concentration
for two values of
$J_{pd}$ = -2.48 eV (triangles) and $J_{pd}$ = -1.5 eV (circles). The
filled (empty) symbols denote disordered (ordered) systems.
}
\label{fig:polarization}
\end{figure}

{\em Spin-polarization:}
Considerable experimental effort has been invested in achieving higher ferromagnetic transition temperatures and higher carrier spin-polarization.
We next discuss the dependence of the carrier spin polarization on the material parameters and impurity concentration.

Our calculation shows that holes in \GaMnAs~ are nearly half-metallic for Mn impurity concentrations $ 0.01 \lesssim x \lesssim 0.05 $ due to the strong exchange coupling, $J_{pd}$= -2.48 eV.
In Fig.~\ref{fig:dos} (a) and (c), deep bound states of Mn clusters pull the majority spin DOS far into the band gap region which fully polarizes the carriers.
The spin-resolved DOS of the ordered system, Fig.~\ref{fig:dos} (b) and (d), also shows nearly complete spin-polarization.
Thus SO coupling has little influence on the carrier spin polarization for systems with strong exchange coupling.

The interplay of SO coupling and disorder is more pronounced in systems where the coupling between the localized spin moment and the carrier spins $|J_{pd}|$ is weaker relative to the SO coupling $\Delta$.
To simulate systems with a smaller $|J_{pd}|/\Delta$, e.g., In$_x$Mn$_{1-x}$Sb, we use GaAs material parameters, i.e., the same SO coupling, but a smaller exchange coupling.
Fig.~\ref{fig:polarization} (a) shows the DOS of GaAs with the magnetic impurity concentration of $x$=0.037 and the exchange coupling $J_{pd}$ = -1.50 eV.
Within our model, $J_{pd}$ = -1.50 eV is too weak for a single impurity to induce bound states but is capable of forming bound states of Mn clusters with non-zero spin-polarization.
For a weaker exchange coupling, SO effects generate considerable weight in the the minority spin DOS, giving a substantial contribution to the polarization for all magnetic impurity concentrations.
We further see, in Fig.~\ref{fig:polarization} (b), the relative enhancement of spin-polarization in disordered systems, compared with the ordered case which grows with increasing impurity concentration.
For a strong exchange coupling, on the other hand, the spin polarization reaches almost 100\% regardless the impurity concentration.

Our results reveal an important aspect of DMS in spintronics applications.
The spin-polarization and mobility of the carriers in a DMS are determined not only by the exchange-coupling but also by the geometrical arrangement of magnetic impurities.
While a stronger exchange-coupling increases the spin-polarization, it also results in more tightly bound states of Mn clusters and, eventually, leads to low mobility of holes.
Our result also indicate that excessive amounts of impurities are unnecessary for optimal spin-transport applications.
With increased impurity concentration, there are more chances of large impurity clusters, which contribute to spin-polarization but degrade the carrier mobility by forming deeper bound states.
To balance the large spin-polarization and the high mobility of itinerant carriers, a delicate modulation of exchange-coupling and impurity randomness is required.

We are grateful to Y.-C.~Chang for helpful discussions about EBOM.
B.L. and R.M.M. were supported by the Office of Naval Research under Grant No. N0014-01-1-1062.
X.C. was supported by the European Commission's Marie Curie International Reintegration Grant No. MIRG-CT-2005-017198 and by Spain's Ministry of Education and Science Ram\'on y Cajal program.


\begin{thebibliography}{22}
\bibitem{footnote1} Due to its nonequilibrium growth process of \GaMnAs~, the carrier concentration is not necessarily the same as the nominal impurity concentration due to charge compensating defects.
We include the effect of charge compensation solely by moving the position of the Fermi energy.
\end{thebibliography}
\end{document}